\pdfoutput=1
\documentclass[aps,prb,twocolumn,superscriptaddress,showpacs, floatfix]{revtex4-1}
\usepackage{amssymb}
\usepackage{graphicx}
\usepackage{amsmath}
\usepackage{amsfonts}
\usepackage{amssymb}
\usepackage{gensymb}
\usepackage[utf8x]{inputenc}
\usepackage{xr}
\usepackage{xcolor}
\DeclareUnicodeCharacter{2212}{-}
\begin{document}
\title{Electron-phonon driven charge density wave in CuTe.}

\author{Marco Campetella}
\email[]{marco.campetella@unisi.it}
\affiliation{Consiglio Nazionale Delle Ricerche (CNR) SPIN, 00133 Rome, Italy}
\affiliation{Dipartimento di Biotecnologie, Chimica e Farmacia, Universit\`a di Siena, Via Aldo Moro 2, Siena, I-53100, Italy}
\author{Giovanni Marini}
\affiliation{Graphene Labs, Fondazione Istituto Italiano di Tecnologia, Via Morego, I-16163 Genova, Italy}
\author{Jianqiang Sky Zhou}
\affiliation{Sorbonne Universit\'e, CNRS, Institut des Nanosciences de Paris, UMR7588, F-75252, Paris, France}
\author{Matteo Calandra}
\email[]{m.calandrabuonaura@unitn.it}
\affiliation{Department of Physics, University of Trento, Via Sommarive 14, 38123 Povo, Italy}
\affiliation{Graphene Labs, Fondazione Istituto Italiano di Tecnologia, Via Morego, I-16163 Genova, Italy}
\affiliation{Sorbonne Universit\'e, CNRS, Institut des Nanosciences de Paris, UMR7588, F-75252, Paris, France}

\begin{abstract}
The compound CuTe (vulcanite) undergoes a quasi one dimensional charge density wave (CDW) at $T< T_{\mathrm{CDW}}=335$ K with a $5\times1\times2$ periodicity. The mechanism at its origin  is  debated. Several theoretical works claimed that semilocal functionals are unable to describe its occurrence and ascribed its formation only to  strong electron-electron interaction. Moreover, the possible role of quantum anharmonicity has not been addressed. Here, by performing quantum anharmonic calculations, we show that  semilocal functionals correctly describe the occurrence of a CDW in CuTe if ultradense electron momentum grids allowing for small electronic temperatures are used. The distortion is driven by the perfect nesting among 1D Fermi surface sheets extending in the $k_y$ direction. Quantum anharmonic effects are important and tend to suppress both the distortion and $T_{\mathrm{CDW}}$. The quantum anharmonic structural minimization of the CDW phase in the generalized gradient approximation leads, however, to distorted Te-Te bond lengths in the low temperature phase that are $21\%$ of the experimental ones at $T=20$ K. This suggests that, even if the electron-electron interaction is not crucial for the mechanism of CDW formation, it is relevant to accurately describe the structural data for the low-T phase. We  assess  the effect of correlation on the CDW by using the DFT+U+V approximation with parameters calculated from first principles. We find that correlation enhances
the Te-Te distortion, T$_{CDW}$ 
and the total energy gain by the distortion.
\end{abstract}

\maketitle

\section{Introduction}

One dimensional (1D) and quasi 1D crystal are  prone to charge density wave (CDW) instabilities due to their low dimensional, often point like, Fermi surfaces and the resulting  divergence in the charge response. This is what is predicted by the Landau-Peierls theory that is characterized by three main features: (i) the transition is second order and manifests itself via a soft phonon going to zero at the transition temperature ($T_{\mathrm{CDW}}$), (ii) the occurrence of an order parameter (the phonon displacement induced by the CDW) that is non-zero only for $T<T_{\mathrm{CDW}}$ and  decreases by increasing temperature until it becomes zero at the transition and, finally, (iii) the opening of a gap in the electronic excitation spectrum whose magnitude should be of the same order of the total energy gain by the CDW distortion. The common belief is that most one dimensional systems are globally well modeled by the Landau-Peierls model.

The Landau-Peierls model is, however, incomplete as it does not account for quantum anharmonicity (i.e. the quantum nature of the ions and the anharmonicity in the ionic potential) that is crucially important for light atoms and in proximity of a second order structural instability. Moreover, it neglects strong electron-electron interaction. Recently the archetypal case of carbyne in vacuum was studied with a variety of density functional theory (DFT) and manybody approaches accounting for non-perturbative quantum anharmonicity and electron-electron interaction\cite{RomaninCarbyne2021}. It was shown that the total energy gain by the distortion is two order of magnitudes smaller (25 meV) than the distortion-induced electronic gap (3 eV) and, most surprisingly, the order parameter increases with increasing temperature for $T<T_{\mathrm{CDW}}$. This  pathology of the carbon chain in vacuum is in part related to the light mass of the carbon atoms and its quantum nature and does not invalidate the applicability of Landau-Peierls theory in a broader spectrum of materials. Still, it implies that there could be remarkable exceptions to this theory, either because quantum anharmonic effects and the electron-electron interaction are crucially important or because the Fermi surface deviates from what expected for a $1D$ system. The applicability of the Landau-Peierls picture to higher dimensional material has been questioned in several works\cite{JohannesPhysRevB.73.205102, JohannesPhysRevB.77.165135}.

\begin{figure}[!htp]
\includegraphics[scale=0.28]
{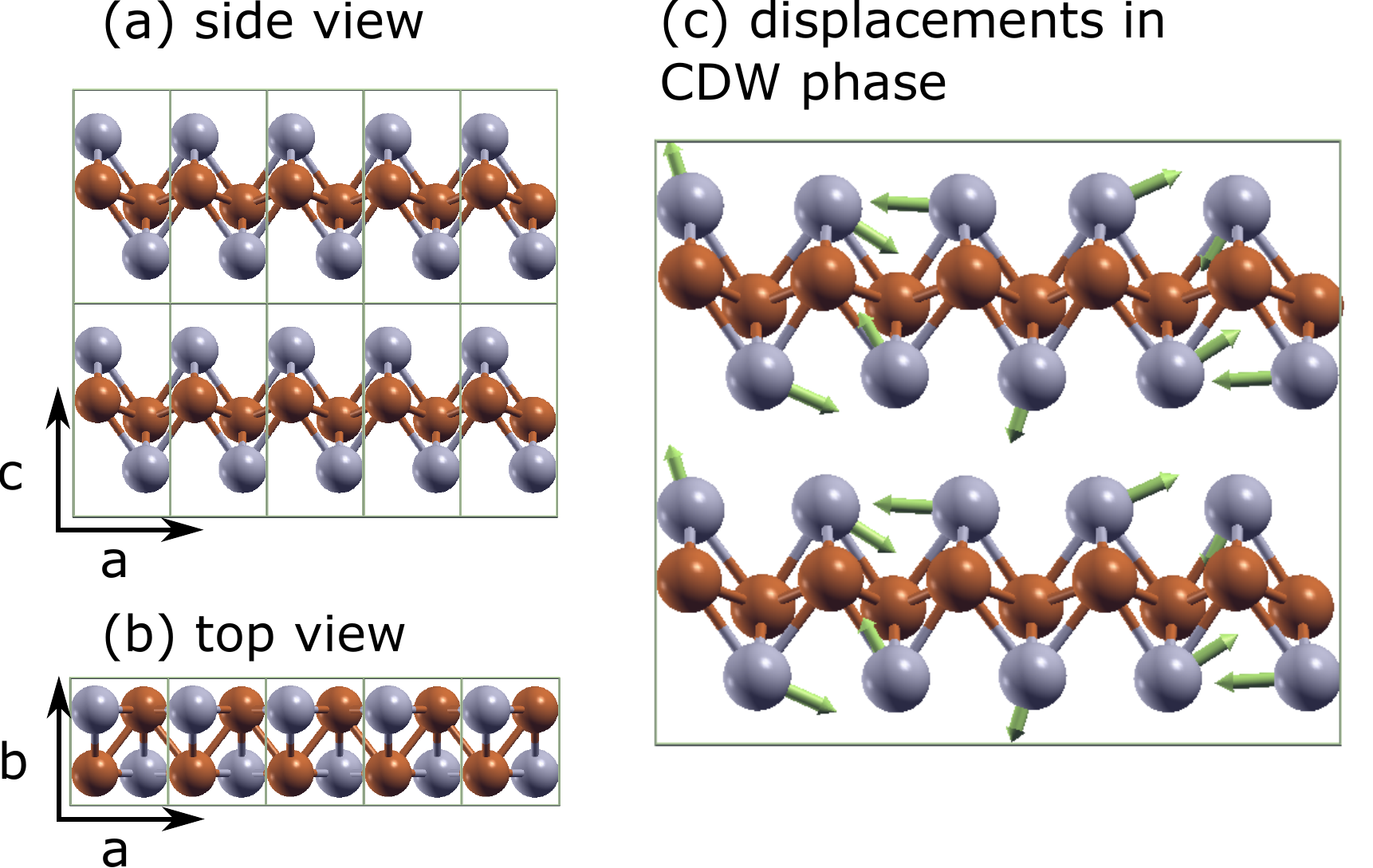}
\caption{Panel (a) and (b): top and side views of the crystal structure of CuTe in the high-T phase on a 5$\times$1$\times$2 cell. Panel (c): side view of the displacements of the Te atoms in the low-T phase (green arrows). The displacements are proportional to  the  phonon eigenvector of the soft mode associated to the CDW. Grey and orange balls represent Te and Cu atoms, respectively.}
\label{fig:struttura}
\end{figure}

Recently, the layered material CuTe (vulcanite) received considerable interest\cite{Wang_APL_2022,zhang2018evidence,kuo2020transport,kim2018large,Kim2019,Salmon-Gamboa2018,CudazzoPhysRevB.104.125101}. In this compound
Te chains run above and below a puckered
copper layer so that each copper atom has a distorted
tetrahedral environment (see Fig.\ref{fig:struttura} and Ref. \onlinecite{stolze2013cute}).
At temperatures lower than T$_{\mathrm{CDW}} = 335$ K,
 CuTe undergoes a $5\times 1\times 2$ CDW \cite{stolze2013cute}. The distortion involves a Te-Te bond alternation with phonon displacements as shown in Fig. \ref{fig:struttura}. The superstructure is
 visible in ARPES data as a  (partial) gapping of the Fermi surface\cite{zhang2018evidence}, the maximum size of the gap being approximately 192 meV\cite{zhang2018evidence}. ARPES data\cite{zhang2018evidence} in the high-T phase show the occurrence of quasi 1D Fermi surface sheets extending along the $k_y$ direction and perfectly nested along $k_x$.
 Resistivity \cite{Wang_APL_2022} and optical\cite{PhysRevB.105.115102} data confirm the quasi 1D character of the CDW as the temperature dependence of the resistivity along the $b$ axis shows the classical behaviour of a metallic system and is not affected by the CDW, while the resistivity along the $a$ axis ( i.e. the CDW direction) displays a marked hump at $T=T_{\mathrm{CDW}}$. Interestingly, the Hall coefficient is enhanced by approximately a factor of two across the CDW transition (larger values of $R_H$ are in the distorted phase) suggesting  a carrier reduction but an incomplete gapping of the Fermi surface in the low-T one\cite{Wang_APL_2022}. The constant pressure specific heat displays a marked jump at the transition albeit with no hysteresis, confirming the second order nature of the transition\cite{kuo2020transport}. 

Thus, this experimental picture seems to point to a second order Landau-Peierls transition mostly due to  the perfect nesting of the quasi 1D Fermi surface sheets.
However, three recent theoretical works \cite{kim2019role,Kim2019,Salmon-Gamboa2018} calculated the harmonic phonon dispersion of the high-T phase of CuTe with semilocal functionals and found no tendency toward CDW (i.e. no imaginary phonon frequencies).
The difficulty in reproducing the occurrence of the CDW with semilocal functionals led some authors to speculate that the CDW in this system is exclusively driven by electron-electron correlation\cite{kim2019role,Kim2019}. Indeed, by performing DFT+U calculations, the authors of Ref. \cite{kim2019role,Kim2019} showed that very large values of U can induce a structural instability comparable with the experimental one. However the considered value for the Hubbard parameter ($U=9$ eV) is extremely large and not calculated ab initio. Furthermore, the role of anharmonicity was not discussed. Recently, a careful study of collective excitations in CuTe \cite{CudazzoPhysRevB.104.125101} pointed out the possible existence of acoustic plasmons, making the study of this compound even more appealing. Finally, CuTe has been reported to support a superconducting state at high pressures \cite{PhysRevB.103.134518}

In this work we investigate the electronic, structural and vibrational properties of CuTe within density functional perturbation theory. We include the effect of non-perturbative quantum anharmonicity by using the Stochastic Self-Consistent Harmonic Approximation \cite{errea2013first,errea2014anharmonic,bianco2017second,monacelli2018pressure,Monacelli_2021}.
We demonstrate that, contrary to what claimed in all published theoretical papers and in agreement with the experimental picture, the CDW is mostly driven by the electron-phonon coupling and Fermi surface nesting with relevant corrections related to quantum anharmonicity. Electron-electron interactions are not negligible but are not the driving force for the CDW transition: they are probably required to accurately describe the structural properties of the low-T phase.

The paper is structured as follows. In Sec. \ref{sec_tec} we give the technical details of the first principles calculations, in Sec. \ref{sec_highT} we address the electronic structure and the mechanism for CDW formation,  in Sec. \ref{sec_lowT} we describe the structural properties of the CDW phase and in Sec. \ref{sec_concl} we draw the main conclusions.

\section{Technical details}\label{sec_tec}
Density-functional theory (DFT) and density functional perturbation theory (DFPT) calculations are carried out using the Quantum ESPRESSO package\cite{giannozzi2009quantum,giannozzi2017advanced}. We use the generalized gradient approximation (GGA) in the Perdew-Burke-Ernzerhof (PBE)\cite{perdew1996generalized} parametrization. 
The experimental measured lattice parameters for bulk CuTe $a = 3.151$ \AA, $b = 4.089$ \AA\ and $c = 6.950$ \AA\ are adopted in all calculations, while we perform structural optimization of internal coordinates. 
We use ultrasoft pseudopotentials\cite{vanderbilt1990soft} and a
50 Ry  plane wave energy cutoff for the kinetic energy (500 Ry for the charge density). 

As phonon dispersion curves in one dimensional materials are extremely sensitive to the k-point sampling and to the electronic temperature ($T_e$) used in the calculation, we perform extremely accurate convergence tests of the phonon frequency at the CDW phonon momentum ${\bf q}_{CDW}=[0.4,0,0.5]$ (square brakets means that the components are given with respect to the basis vectors of the reciprocal lattice).
In more details, 
the harmonic phonon dispersion is calculated using  $\Gamma$ centered k-points meshes. We considered grids of the kind $k_x\times16\times 4$ with $k_x$ values up to $150$. We then calculate the phonon frequency for each  mesh as a function of the Fermi temperature used in the calculations. The results of these calculations are
explained in more details in Sec. \ref{sec_highT}. At the end of these tests we adopted an  80$\times$16$\times$4 electron-momentum grid in the 1$\times$1$\times$1 cell  and an electronic temperature $T_e=200$ K (Fermi Dirac smearing).  
When using supercells, the k-points meshes are then rescaled according to the size of the supercells (e.g., we use a 8$\times$16$\times$2 k-points mesh on a 10$\times$1$\times$2 cell and a 16$\times$16$\times$4 k-points mesh on a 5$\times$1$\times$1 cell). 

The quantum anharmonic calculation is performed with the Stochastic Self Consistent Harmonic Approximation (SSCHA)\cite{errea2013first,errea2014anharmonic,bianco2017second,monacelli2018pressure,Monacelli_2021}.
The SSCHA is a stochastic variational technique that allows to access the non-perturbative quantum anharmonic free energy and its Hessian with respect to the atomic positions  \cite{bianco2017second} (\textit{i.e.}, the phonon spectrum). The SSCHA technique requires the evaluation of forces in supercells with atoms displaced from their equilibrium positions following a suitably chosen Gaussian distribution. The forces can be calculated by using any force engine. In this work 
we used DFT with the PBE functional for the force calculation. 
We calculate the forces using the Quantum ESPRESSO package and supercells ranging from $5\times 1\times2$ to $10\times 1\times2$. In a  
  10$\times$1$\times$2 supercell (80 atoms) of the high-T phase structure the number of DFT force calculations needed to converge the free energy is
of the order of $800$, while approximately $2000$ forces are needed to converge the free energy Hessian at $T = 0$~K. The computational effort is substantial given the dense electron-momentum grids.

We determine the nature and the critical temperature $T_{\mathrm{CDW}}$ of the CDW transition by monitoring the positional free energy Hessian (second derivative of the free energy with respect to the atomic positions)\cite{bianco2017second}, as dictated by Landau theory of phase transitions. 

\section{High-T phase}\label{sec_highT}

\begin{figure*}[!htp]
\includegraphics[scale=0.34]
{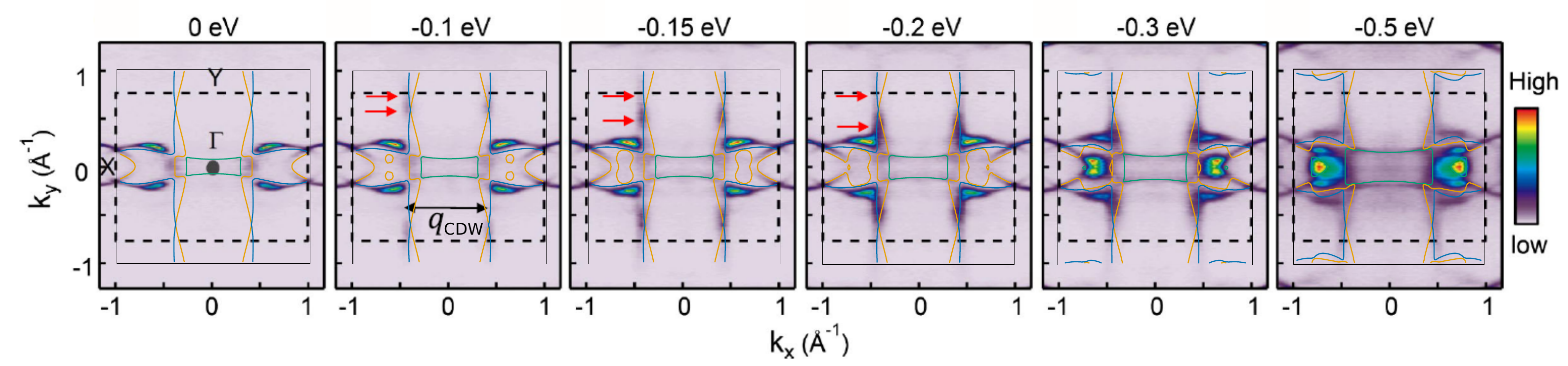}
\caption{Constant energy cuts from $E_F$ (0 eV) to -0.5 eV from $E_F$ (the value of the constant energy with respect to $E_F$ is shown on the top of each panel) in the $(k_x,k_y)$ plane for $k_z=0$. Experimental ARPES intensity data are from Ref. \onlinecite{zhang2018evidence}.
 Theory are the continuous yellow, green and blue lines.}
\label{fig:fermi_surface}
\end{figure*}

We first calculate the electronic structure of the high-T phase and compare the Fermi surface with that measured in ARPES (see Fig. \ref{fig:fermi_surface}).
Each panel refers to constant energy cuts from $E_F$ to -0.5 eV from $E_F$ (the value of the constant energy with respect to $E_F$ is shown on the top of each panel) in the $(k_x,k_y)$ plane and for $k_z=0$. Experimental ARPES data from Ref. \onlinecite{zhang2018evidence}
are also included for reference.

Globally the agreement between the experimental and measured constant energy scans is  excellent.  We are able to recover both the pockets extending along the $k_x$ direction  and the quasi-1D line segments along the $k_y$ direction. These last dispersionless bands extending only along the k$_y$ direction are clear fingerprints of the 1D physics in vulcanite.

\begin{figure}[!htp]
\includegraphics[scale=0.28]
{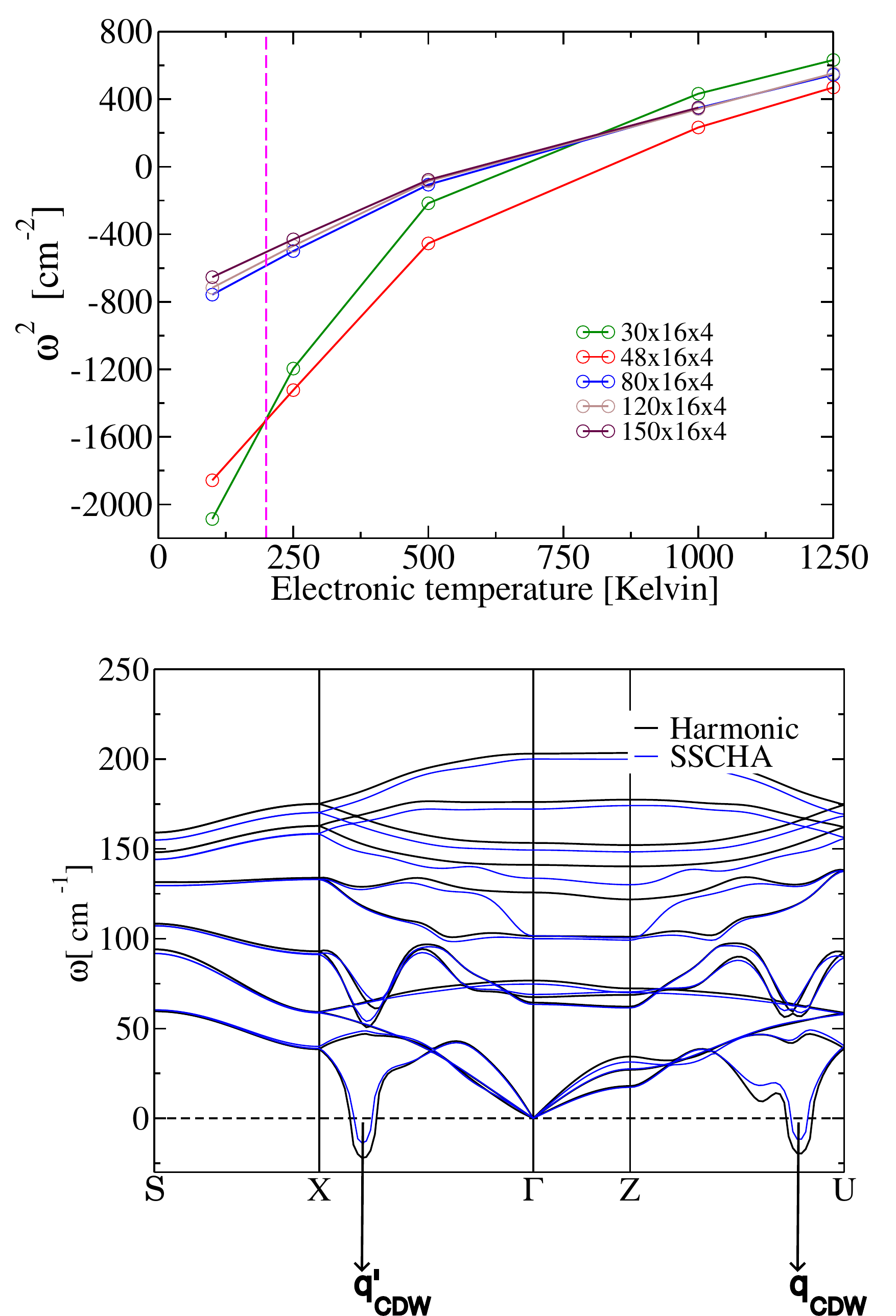}
\caption{Top Panel : convergence of the lowest phonon frequency at the q$_{\mathrm{CDW}}$ wave vector with respect to the k-grid (k$_x\times$16$\times$4) and smearing temperature at harmonic level. The magenta vertical dashed line show the electronic temperature adopted  in this work by using an $80\times 16\times 4$ electron momentum grid. Imaginary phonon frequencies are represented as negative values. Bottom Panel : The harmonic (black line) and quantum anharmonic (blue line) phonon bands (square root of the eigenvalues of the free energy Hessian divided by the masses) calculated for the high-T phase at $T = 0$ K.}
\label{fig:w_T0}
\end{figure}

The sharpness of these 1D Fermi surface portions suggests that a remarkably dense k-points mesh along the $k_x$ direction may be required in order to correctly sample their contribution to the phonon dispersion at phonon momentum ${\bf q}={\bf q}_{\mathrm{CDW}}$. 
We explicitly verified this point by performing careful convergence of the lowest energy phonon frequency at ${\bf q}={\bf q}_{\mathrm{CDW}}$ as a function of $k_x$ points and Fermi-Dirac electronic temperature T$_e$. The results are shown in Fig. \ref{fig:w_T0} (top)  and unambiguously show that grids having $k_x\approx 80$ and electronic temperatures comparable to T$_{CDW}$ must be used to see the CDW. By adopting an electronic temperature $T_e=200$ K and a k-point mesh of $80\times 16\times 4$ we find  converged results. As it can be seen, the lowest phonon frequency at ${\bf q}={\bf q}_{\mathrm{CDW}}$ is imaginary and not positive as it has been reported in all published theoretical papers in the field\cite{kim2018large,kim2019role,Salmon-Gamboa2018}. 
In these works, the difficulty in performing Brillouin zone sampling for CuTe  has been completely overlooked. Much coarser grids, such as $30\times16\times 4$, and, most likely, larger electronic temperatures have been used.  We point out that the technical details reported in Refs. \cite{kim2019role,Kim2019,Salmon-Gamboa2018} are incomplete and the calculations are not reproducible (as an example in Refs. \cite{kim2019role,Kim2019} the value of the electronic temperature is not reported).

From Fig. \ref{fig:w_T0}, it is also clear that by using the PBE semilocal functional and simply increasing the electronic temperature, i.e. neglecting quantum anharmonicity, the CDW critical temperature is in the range $T_{\mathrm{CDW}}=700-750$ K. This is only a factor of two higher than the experimental one, suggesting that Fermi surface nesting is an important effect in this system.

The CuTe harmonic phonon dispersion is reported in Fig. \ref{fig:w_T0} (bottom panel). As it can be seen there are two sharp dynamical instabilities corresponding to the modulations ${\bf q}_{\mathrm{CDW}}=[0.4,0,0.5]$ and
${\bf q}_{\mathrm{CDW}}^{\prime}=[0.4,0,0.0]$. The planar instability at ${\bf q}_{\mathrm{CDW}}^{\prime}$ leads to slightly more unstable phonons. However small changes in the simulations details (structural parameters, functional used,...) lead to a more unstable mode at ${\bf q}_{\mathrm{CDW}}$. These two instabilities are then almost degenerate. The local character in momentum space of the instability points at a crucial role of the Fermi surface. 

\begin{figure}[!h]
\includegraphics[scale=0.28]{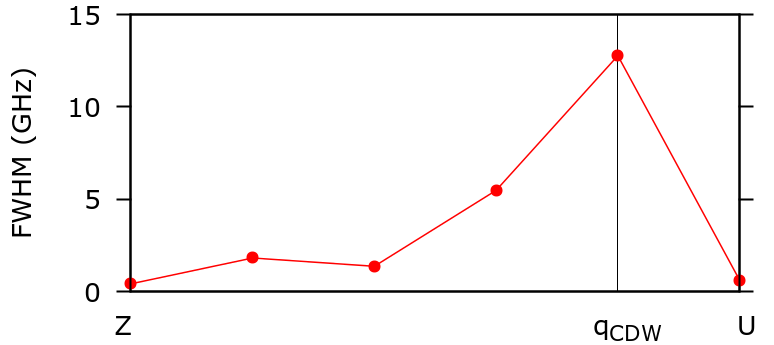}
\caption{Phonon linewidth of the low energy mode as a function of phonon momenta along the ZU direction. }
\label{fig:gamma}
\end{figure}

In order to confirm this point we calculate the electron-phonon contribution to the phonon linewidth (FWMH), namely
\begin{equation}
    \gamma_{{\bf q}\nu}=\frac{4 \pi\omega_{{\bf q}\nu}}{N_k}\sum_{{\bf k},n,m}\left|g_{{\bf k} n,{\bf k}+{\bf q} m}^\nu\right|^2\delta(\epsilon_{{\bf k} n}-E_F)\delta(\epsilon_{{\bf k}+{\bf q} m}-E_F)
\end{equation}
where $\omega_{{\bf q}\nu}$ are the harmonic phonon frequencies, $\epsilon_{{\bf k}n}$ are the Kohn-Sham energy bands, $E_F$ is the Fermi level and $g_{{\bf k} n,{\bf k}+{\bf q} m}^\nu$ is the electron-phonon matrix element. We calculate $\gamma_{{\bf q}\nu}$ for the lowest energy phonon mode along the ZU direction. The results are shown in Fig. \ref{fig:gamma} and shows a strong enhancement of the phonon linewidth at the CDW wavevector mostly due to Fermi surface nesting. At the harmonic level and by using the PBE functional the instability is then electron-phonon driven.

The phonon patterns connected with these two instabilities are very similar in the CuTe \textit{ab}-plane.
The only difference is that the distortion of momentum  ${\bf q}={\bf q}_{\mathrm{CDW}}$ shifts two parallel CuTe planes in antiphase.
The calculation of the energy gain obtained by displacing the ions along the directions of the imaginary phonon mode is approximately $1.29$ meV per Cu atom in both cases.

The occurrence of imaginary phonon frequencies at the harmonic level is, however, not enough to demonstrate the presence of a CDW as quantum-anharmonic terms in the potential could remove the instability. In order to explore this possibility, 
 we investigate quantum anharmonic effects within the Stochastic Self-Consistent Harmonic Approximation (SSCHA)\cite{errea2013first,bianco2017second,errea2014anharmonic,monacelli2018pressure,zhou2020anharmonicity} that has been proven to be very effective in describing anharmonic quantum effects in a plethora of different systems\cite{aseginolaza2019phonon,bianco2019quantum,errea2020quantum}.

The quantum anharmonic phonon dispersion is obtained within the SSCHA by calculating the positional free energy ($F$) Hessian as a function of temperature. We define the temperature dependent dynamical matrix as:
\begin{equation}
    \textbf{\textit{D}} = \textbf{M}^{-\frac{1}{2}}\frac{\partial^2 F}{\partial \textbf{R}^2} \Bigg| _{\textbf{R}_{eq}} \textbf{M}^{-\frac{1}{2}}
\end{equation}
where $\textbf{M}$ is the matrix of the ionic masses $M_a$ with $M_{ab}$ = $\delta_{ab} M_a$ and $\textbf{R}$ is a cumulative variable for all the ionic positions (see Ref. \onlinecite{bianco2017second} for a detailed explanation). By  Fourier transforming the matrix $\textbf{\textit{D}}$ and  by diagonalizing it, we obtain as eigenvalues the squared quantum anharmonic phonon frequencies.

\begin{figure}[!htp]
\includegraphics[scale=0.28]
{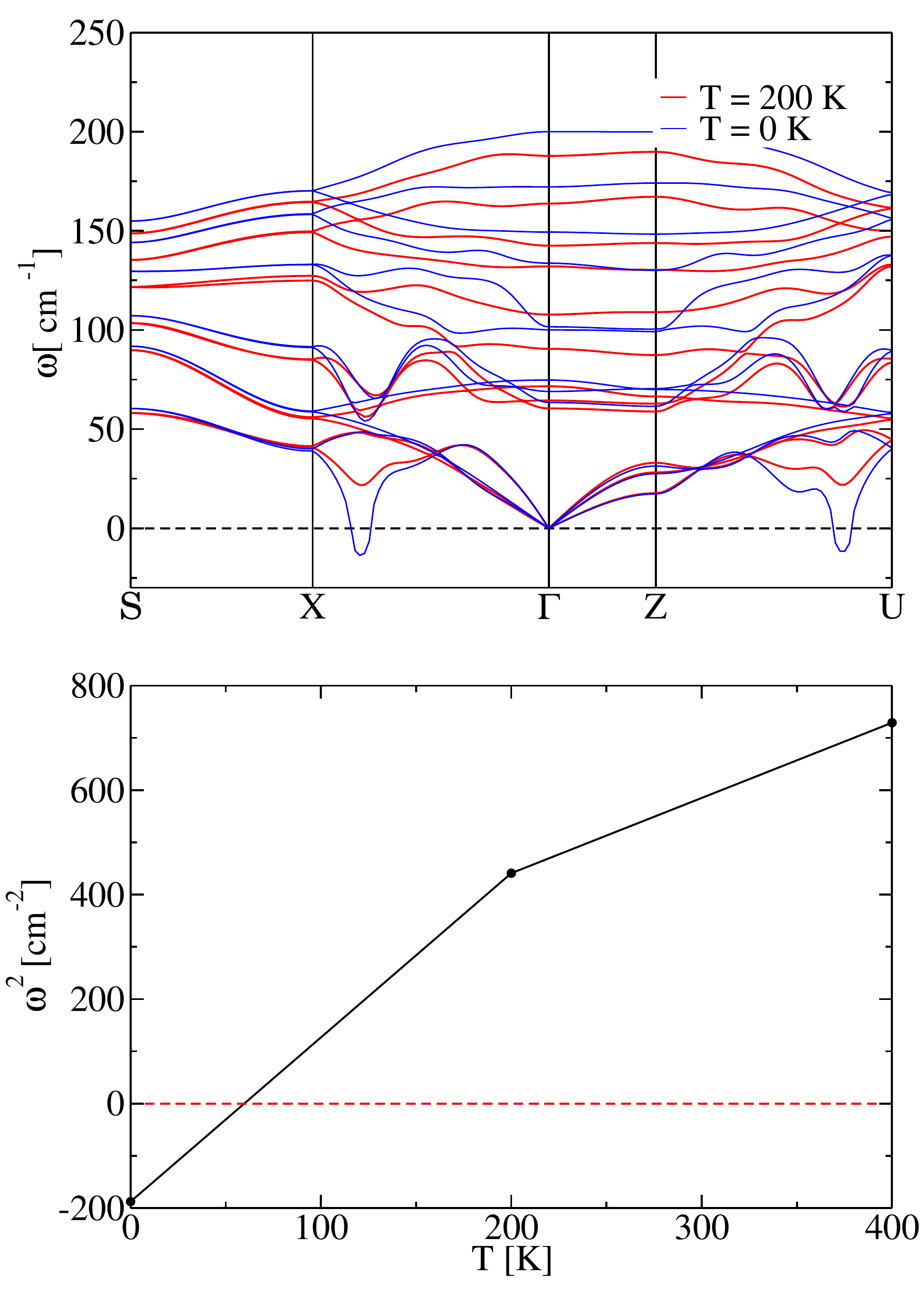}
\caption{Top panel : anharmonic phonon dispersion (square root of the eigenvalues of the free energy Hessian divided by the masses) computed at 0 K and 200 K respectively. Bottom Panel : square of the lowest phonon frequency computed at phonon momentum ${\bf q}={\bf q}_{\mathrm{CDW}}$ as a function of T. }
\label{fig:w2_vs_T_free}
\end{figure}

We perform the SSCHA calculation on a 10$\times$1$\times$2 supercell. The results are shown in Fig. \ref{fig:w_T0} ($T=0$, bottom panel) and in Fig. \ref{fig:w2_vs_T_free} (top panel) as a function of temperature. At $T=0$ the main effect of quantum anharmonicity is an hardening of the CDW mode. However, the mode still remains imaginary signalling that at $T=0$ quantum anharmonicitiy does not remove the CDW.

The temperature dependence of the quantum anharmonic phonon dispersion  is shown in Fig. \ref{fig:w2_vs_T_free} (top panel). At $T=200$ K the phonon dispersion does not display any dynamical instability, meaning that the calculation is already in the undistorted high-T phase. 
By plotting the square of the lowest phonon frequency as a function of temperature in Fig. \ref{fig:w2_vs_T_free} (bottom panel) we estimate $T_{\mathrm{CDW}}\approx 60$ K. This critical temperature is approximately $5.6$ times smaller than the real one. As the transition occurs only via a change in the quantum free energy Hessian that becomes negative at the transition along the CDW pattern, we find that in our calculation the transition is purely second order, in agreement with experimental data \cite{kuo2020transport}.

Two effects may be at the origin of the underestimation of $T_{\mathrm{CDW}}$. The first one is that the supercell used in the calculation could be too small. However, we have carefully monitored the value of the phonon frequency at ${\bf q}={\bf q}_{\mathrm{CDW}}$ and ${\bf q}={\bf q}_{\mathrm{CDW}}^{\prime}$ for supercells of sizes $5\times 1\times 1$, $5\times 1\times 2$, $10\times 1\times 2$
finding that the quantum anharmonic phonon frequency varies less than $1$ cm$^{-1}$. This excludes that this reduced $T_{\mathrm{CDW}}$ is due to a finite supercell effect.

The second and most probable reason causing the underestimation of $T_{\mathrm{CDW}}$ is the treatment of the exchange and correlation used. In order to better understand this point we examine more in details the low temperature phase.

\section{Low temperature CDW phase.}\label{sec_lowT}
 
 In order to study the structural and electronic properties of the CDW phase, we consider two supercells, the $5\times 1\times 1$ and the $5\times1\times2$, corresponding to instabilities at ${\bf q}={\bf q}_{\mathrm{CDW}}^{\prime}$ and
 ${\bf q}={\bf q}_{\mathrm{CDW}}$, respectively.
 We first displace the atoms along the unstable phonon patterns and then perform structural optimization (we minimize the classical Born-Oppenheimer forces). The results of the optimization are shown in Tab. \ref{tab1} . As it can be seen, the structural distortion of the Te atoms is in good agreement with experiments at $T=20$ K, although the distortion is somewhat underestimated. Both the $5\times 1\times 1$ and the $5\times1\times2$ give comparable 1D distortion. 
 
 The fact that, as we have seen, quantum anharmonic effects are important in this system, as they reduce $T_{\mathrm{CDW}}$ more than a factor of $10$ with respect to the harmonic calculation, suggests that the inclusion of quantum anharmonicity will reduce the distortion. As the quantum anharmonic minimization in this system is very expensive due to the very dense mesh needed, we perform the quantum anharmonic structural optimization with the SSCHA only in the $5\times 1\times 1$ supercell. This is justified as we know that the two supercells lead to practically identical distortion of the Te-Te bond along the CDW direction.
 
 The results of the quantum anharmonic minimization are again shown in Tab. \ref{tab1}. As expected the distortion is substantially reduced and the quantum anharmonic distortion is approximately $41\%$ ($0.21\% $) of the experimental one at $T=295$ K ($T=20 $K).
 As in low dimensional systems it is well known that the exchange interaction is not completely screened and semilocal functional usually underestimate the distortion\cite{RomaninCarbyne2021}, this has to be somewhat expected.

\begin{table*}
    \centering
    \begin{tabular}{|c|c|c|c|c|}
    \hline
  CDW periodicity & Te-Te min. distance (\AA) & Te-Te max. distance (\AA) & $\Delta_{\rm Te}$ (\AA) & $\Delta_F$ (meV/f.u.)\\
 \hline
     \hline
          5$\times$1$\times$1 (classical ions) & 3.084 & 3.222 & 0.138 & 1.47 \\ 
    \hline
     5$\times$1$\times$2 (classical ions) & 3.07 & 3.21 & 0.14 & 1.29 \\
    \hline
    5$\times$1$\times$1 (quantum  ions SSCHA) & 3.109 & 3.196 & 0.087 & 0.47 \\ 
 \hline

    measured ( $T = 295$ K; Ref.\onlinecite{stolze2013cute}) & 3.05 & 3.26 & 0.21 & // \\
              \hline
         
    measured ( $T = 20$ K; Ref.\onlinecite{stolze2013cute}) & 2.95 & 3.32 & 0.37 & // \\ \hline    
    \end{tabular}
    \caption{Comparison among first principles structural parameters for 5$\times$1$\times$1 within the SSCHA, 5$\times$1$\times$1 and 5$\times$1$\times$2 within the harmonic approximation and experimental CDW distortion patterns. The quantity $\Delta_{\rm Te}$ is the difference among the maximal and minimal Te-Te distances. $\Delta F$ represents the total (electronic plus vibrational) free energy gain per formula unit due to the CDW distortion.}
    \label{tab1}
\end{table*}

Finally, for completeness, we address the pseudogap feature detected in ARPES\cite{zhang2018evidence} in the CDW phase. Previous calculations already showed that this feature can be fairly well reproduced if the distortion is large enough \cite{kim2018large,Kim2019}. As it is typical for a Peierls distortion, the magnitude of the gap opening is linearly related to the CDW distortion.
This means that, as the magnitude of the distortion depends on the exchange and correlation approximation used in the calculation, the size of the pseudogap also will.

We then consider the experimental distorted structure on a $5\times1\times 2$ supercell, calculate the electronic structure and unfold it \cite{bandsup1_PhysRevB.89.041407, bandsup2_PhysRevB.91.041116} on the CuTe unit cell. A finite Lorentzian linewidth of 20 meV is added to the theoretical unfolded band structure in order to simulated the experimental broadening. The comparison with ARPES data from Ref. \cite{zhang2018evidence} is also shown. Our calculations reproduce the opening of the CDW with a pseudogap that is of the same magnitude of the experimental one. Small differences occur on the exact value of the experimental gap that are probably due in part to the ARPES matrix element, not explicitly considered in our calculation.
\begin{figure}[!htp]
\includegraphics[scale=0.14]
{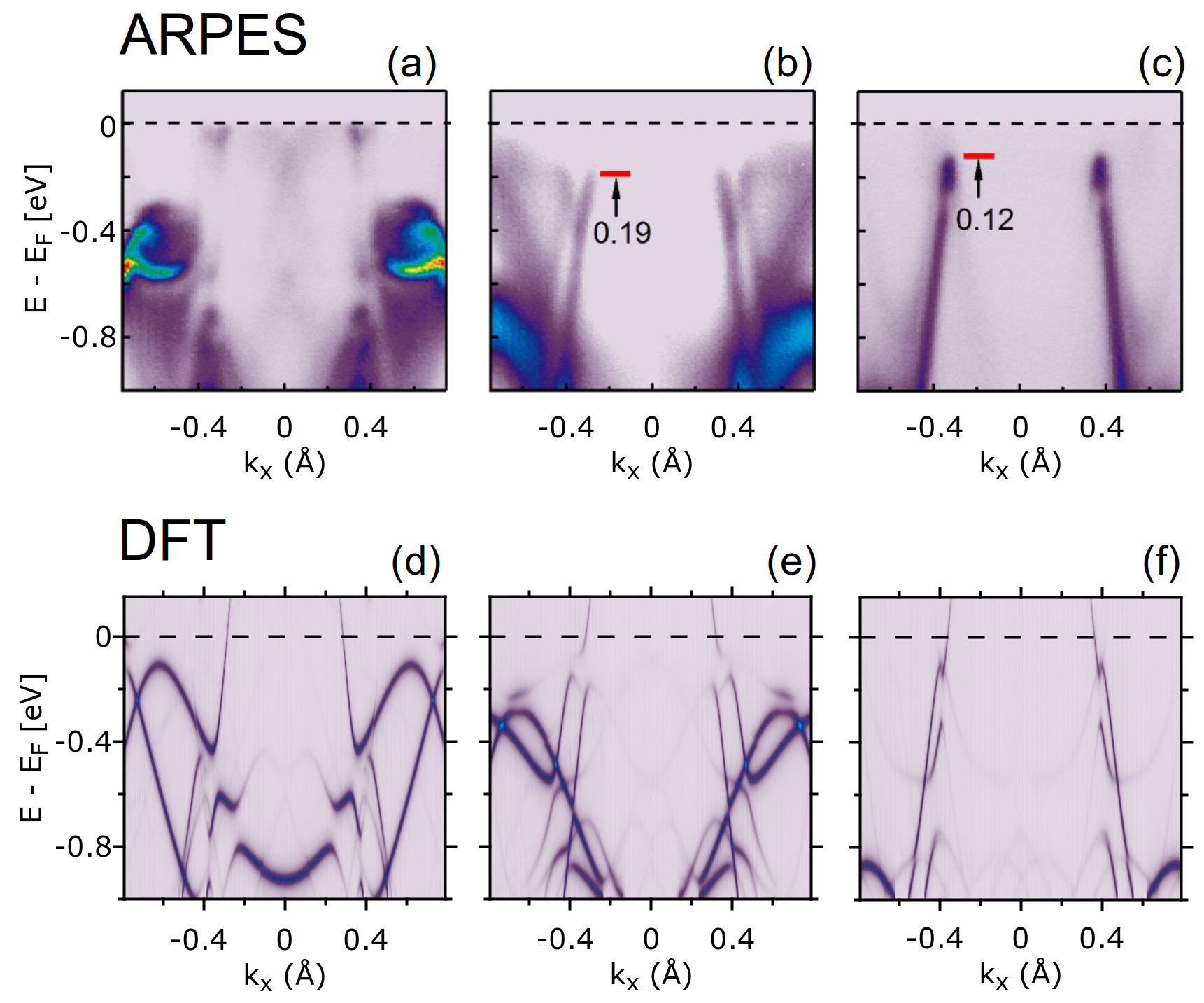}
\caption{Comparison between experimental ARPES spectrum at $k_y = 0, 0.3, 0.55~$ \AA$^{-1}$ (panels (a),(b),(c) respectively, taken from Ref. \onlinecite{zhang2018evidence}) and first principles unfolded band spectrum for the 5$\times$1$\times$2 distortion pattern (panels (d), (e), (f), respectively). The experimental amplitude of the CDW distortion has been used in this calculation (see Tab.\ref{tab1}).}
\label{fig:unfolding}
\end{figure}

\begin{figure*}[!htp]
\includegraphics[scale=0.34]
{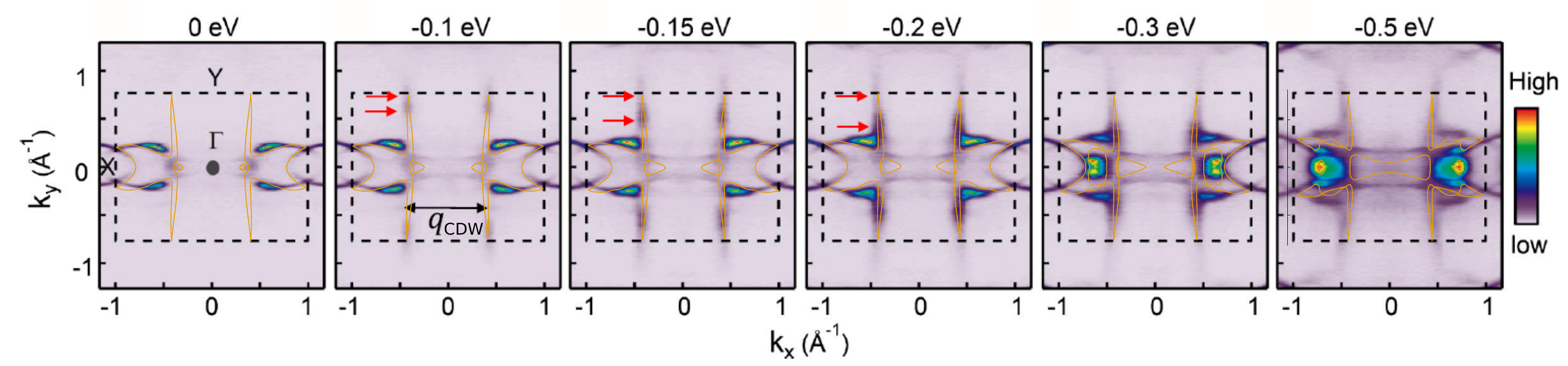}
\caption{Constant energy cuts from $E_F$ (0 eV) to -0.5 eV from $E_F$ (the value of the constant energy with respect to $E_F$ is shown on the top of each panel) in the $(k_x,k_y)$ plane for $k_z=0$ using calculated first principles Hubbard parameters of Tab.\ref{tab2}. 
 Theory are the continuous yellow lines.}
\label{fig:fermi_hubbard}
\end{figure*}

\section{Estimation of correlation effects via  DFT+U+V}\label{sec_DFTUV}

In order to account for correlation effects 
on the electronic structure and the structural properties on equal footing,  we model the system in the DFT+U+V formalism within the rotationally invariant scheme first proposed by Dudarev $et al.$ in Ref. \onlinecite{PhysRevB.57.1505}. Following Ref. \onlinecite{Leiria_Campo_2010}, the DFT energy functional, $E_{DFT}$, is corrected to include on-site and inter-atomic interactions, by adding the term

\begin{equation} 
\begin{gathered}  
E_{UV} = \sum_{I}\frac{U^I}{2} \textrm{Tr}[\mathbf{n}^{II\sigma} (\mathbf{1} - \mathbf{n}^{II\sigma})] \\
- \sum^*_{I,J,\sigma} \frac{V^{IJ}}{2} \textrm{Tr}[\mathbf{n}^{IJ\sigma}\mathbf{n}^{JI\sigma}]
\end{gathered}
\end{equation}

where $I$ and $J$ represent atomic sites, the star in the sum operator denotes that for each atom $I$, $J$ covers all its neighbors up to a given distance, while the on-site parameter U${^I}$, the inter-site V$^{I,J}$ and the occupation matrix $\mathbf{n}^{IJ\sigma}$ are defined as in Ref. \onlinecite{Leiria_Campo_2010}.

The new total energy E$_{DFT+U+V}$ is written as 

\begin{equation} 
E_{DFT+U+V} = E_{DFT} + E_{UV}
\end{equation}

The  on-site and intersite parameter $U^I$ and $V^{IJ}$ parameters are calculated from first principles, using the linear response method introduced by Timrov $et$ $al.$ in Refs. \onlinecite{PhysRevB.98.085127,PhysRevB.103.045141}.

We use the atomic wavefunctions (3$d$ for Cu and 5$p$ for Te) read from the pseudopotentials to build the Hubbard projectors. In the calculation, all the neighboring atoms up to the fourth shell were considered. A  8$\times$4$\times$1 momenta grid was necessary to converge the U and V values within 0.1 eV. 
The calculated inter and on-site Hubbard values for CuTe in the normal phase are reported in Tab. \ref{tab2}.

\begin{table}
    \centering
    \begin{tabular}{|c|c|c|c|c|}
    \hline
 Atom 1 & Atom 2  & shell  & distance (\AA)~& V$^{IJ}$ (eV)\\
     \hline
     Cu  &   Cu  &  0 & 0  & 16.71 \\
     Te  &   Te  &  0 & 0 & 4.32 \\
     Cu  &   Te  &  1 & 2.6 & 0.08  \\
     Cu  &   Cu  &  2 & 2.66 & -0.12 \\
     Cu  &   Te  &  3 & 2.661 & -0.09  \\   
     Te  &   Te  &  4 & 3.15 & 0.97 \\
    \hline
    \end{tabular}
    \caption{Calculated $U^{I}$ and $V^{IJ}$ values for the generalized Hubbard model in CuTe for the first five neighbor shells (shell 0 corresponds to the on-site term) and corresponding distances. Here, the notation $V^{II}$ = $U^{I}$ is employed.}
    \label{tab2}
\end{table}

We find large on-site repulsion parameter of 16.71 eV and 4.32 eV for Cu(3$d$) and Te(5$p$) sites, respectively. Furthermore, we observe that interatomic Cu-Te interactions are negligible, while a sizable first-neighbor Te-Te repulsive interaction (0.97 eV) exists. The inclusion of Hubbard parameters importantly modifies the electronic structure, resulting in the Fermi surface shown in Fig.\ref{fig:fermi_hubbard} (top panel, yellow lines). By looking at the comparison between the ARPES and the Fermi surface predicted by first principles calculations employing DFT+U+V, we conclude that the first principles on-site and inter-site parameters are not substantially improving the agreement between the theory and the experiment, especially in regard to the electron pocket around the $\Gamma$ point.

\begin{figure}[!htp]
\includegraphics[scale=0.31]
{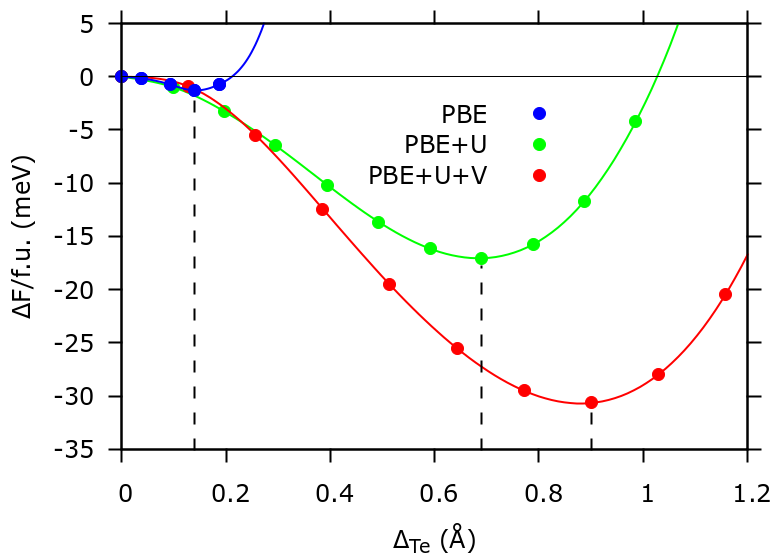}

\caption{Energy gain along the Te-Te maximum dimerization, $\Delta_{Te}$, with (filled green dots) and without (filled blue dots) the inclusion of the Hubbard U on the Cu($3d$) orbitals ( = 10 eV). The lines are guides to the eye. }
\label{fig:enegain}
\end{figure}

Finally, we calculate the energy gain in  the charge-density wave phase with respect to the normal state with the inclusion of Hubbard parameters, and compare the results to the predictions given by PBE. The results are depicted in Fig. \ref{fig:enegain}. We find that the inclusion of correlation effects enhances the CDW energy gain by more than one order of magnitude, i.e. from 1.29 meV /f.u. in PBE to 32 meV/f.u. if both inter- and on-site parameters are included in the calculation, while we obtain an energy gain of 17 meV/.f.u. if only on-site terms on Cu and Te are included in the calculation. Correspondingly, the predicted structural distortion due to the charge-density wave is notably enhanced, with a maximum Te-Te dimerization $\Delta_{\rm Te}$ of the order of 0.9 \AA~,  overestimating the measured values of Ref.\onlinecite{stolze2013cute} of a factor $\approx 2.4$ at $T=20$ K.  Moreover the free energy versus $\Delta$ profile becomes even more anharmonic, suggesting both an increase of T$_{CDW}$ at the harmonic level as well as an enhancement of quantum anharmonic effects.

As it was already clear at the PBE level, the charge density wave temperature is the result of a delicate
compensation among the electron-phonon interaction (enhancing the tendency towards CDW) and anharmonicity (suppressing the CDW). Both effects are substantially enhanced by correlation effects and both effects are crucial and comparable in order. Within DFT+U+V at the harmonic level, we do indeed estimate T$_{\rm CDW}$ as being as large as 6000 K, in stark disagreement with experiments, signalling once more the need of including anharmonicity to obtain results in better agreement with experiments.

\section{Conclusion }\label{sec_concl}

In this work, by performing non perturbative quantum-anharmonic calculations, we studied the CDW formation in CuTe.
Contrary to all existing theoretical calculations in literature \cite{kim2019role,Kim2019,Salmon-Gamboa2018}, we find that semilocal functionals correctly describe the occurrence of CDW in this system.
Previous calculations where unable to describe the CDW instability  most likely due to the use of a too large electronic temperature.

We find that the CDW is due to the almost perfect nesting among the quasi 1D Fermi surface sheets extending along the $k_y$ direction resulting in a large electron-phonon interaction and a consequent phonon softening. Quantum anharmonicity reduces this softening but does not suppress the CDW at $T=0$.
Quantum anharmonic effects reduce the $T_{\mathrm{CDW}}$ by  a factor of 10 with respect to the harmonic estimate based on the electronic temperature only.

The calculated $T_{\mathrm{CDW}}\approx 60$ K, resulting from the combined effect of the electron-phonon interaction and anharmonicity, underestimates the experimental one by a factor $\approx 5.6$. Similarly, the quantum anharmonic structural minimization of the CDW phase leads to distorted Te-Te bond lengths in the low temperature phase that are $40\%$ smaller than the experimental ones. These two underestimations are related and suggest that, even if the electron-electron interaction is not crucial for the mechanism of CDW formation, it is relevant to accurately describe the structural data for the low-T phase.

In order to validate this statement we employ the DFT+U+V approximation with on-site and off-site Hubbard parameters calculated ab initio. Within this approximation, the CDW distortion is strongly enhanced and overestimate the experimental one by a factor 5.6. At the harmonic level T$_{CDW}\approx 6000 $K, approximately $20$ times larger than the experimental value. However, anharmonic effects also becomes substantially larger, underlying once more the need of including quantum anharmonic effects to obtain results in better agreement with experiments.

\section{Acknowledgements}

 Co-funded by the European Union-NextGenerationEU, ICSC – Centro Nazionale di Ricerca in HPC, Big Data and Quantum Computing. Views and opinions expressed are however those of the author(s) only and do not necessarily reflect those of the European Union or the European Research Council. Neither the European Union nor the granting authority can be held responsible for them.
We acknowledge the PRACE and CINECA award under the ISCRA initiative, for the availability of high performance computing resources and support. We acknowledge support from Seal of Excellence (SoE) fellowship promoted by University of Siena

\bibliographystyle{elsarticle-num.bst}
\bibliography{reference.bib}


\end{document}